\documentclass[prb,aps,showpacs,floatfix,floats,nobalancelastpage,twocolumn]{revtex4}
\usepackage{graphicx}
\usepackage{latexsym,amsmath,amssymb,bm,euscript}
\usepackage{color}
\usepackage{dcolumn}
\usepackage{bm}
\usepackage{epstopdf}
\usepackage{times}
\usepackage{multirow}

\def\ra{\rangle}
\def\la{\langle}
\def\up{\uparrow}
\def\dn{\downarrow}
\def\Hc{{\rm H.c.}}

\begin{document}

\title{Insulating phases of electrons on a zigzag strip in the orbital magnetic field}
\author{Hsin-Hua Lai}
\affiliation{Department of Physics, California Institute of Technology, Pasadena, California 91125, USA}
\author{Olexei I. Motrunich}
\affiliation{Department of Physics, California Institute of Technology, Pasadena, California 91125, USA}
\date{\today}

\begin{abstract}
We consider electrons on a two-leg triangular ladder at half-filling and in an orbital magnetic field.  In a two-band regime in the absence of the field, the electronic system remains conducting for weak interactions since there is no four-fermion Umklapp term.  We find that in the presence of the orbital field there is a four-fermion Umklapp and it is always relevant for repulsive interactions.  Thus in this special ladder, the combination of the orbital magnetic field and interactions provides a mechanism to drive metal-insulator transition already at weak coupling.  We discuss properties of the possible resulting phases C0S2 and various C0S1 and C0S0.
\end{abstract}
\maketitle

\section{Introduction}
This paper complements our earlier work Ref.~\onlinecite{SBMZeeman} on the effects of Zeeman field on a Spin Bose-Metal (SBM) phase\cite{Sheng09} (the reader is referred to Refs.~\onlinecite{SBMZeeman, Sheng09} for general introduction).  Here we consider the orbital magnetic field on the electronic two-leg triangular ladder.

Previous studies of ladders with orbital field were done on a square 2-leg case and mainly focused on generic density (see Refs.~\onlinecite{Roux07, Narozhny2005, Carr2006, Orignac2001} and citations therein), while the triangular 2-leg case has not been considered so far.
In the context of Mott insulators at half-filling, microscopic orbital fields were shown to give rise to interesting scalar chirality terms operating on triangles in the effective spin Hamiltonian.\cite{Rokhsar90, Sen95, magnetoorb, Katsura2010} On the other hand, it was also argued\cite{Bulaevskii08, Al-Hassanieh2009, Khomskii10} that if a Mott insulator develops a noncoplanar magnetic order with nontrivial chiralities, this can imply spontaneous orbital electronic currents.

In this paper, we focus on the simplest ladder model with triangles, the zigzag strip, and discuss instabilities due to existence of orbital magnetic field and properties of the resulting phases.  Our main findings are presented as follows.  In Sec.~\ref{weakcoupling:orbital}, we determine the electron dispersion in the orbital field and perform weak coupling renormalization group (RG) analysis in a two-band regime.\cite{Sheng09, Lai10, Louis01, Fabrizio96}  Unlike the case with no field, we find that there is a four-fermion Umklapp interaction which is always relevant for repulsively interacting electrons and provides a mechanism to drive the metal-insulator transition.  This Umklapp gaps out all charge modes and produces a C0S2 state.  In Sec.~\ref{observables} we describe physical observables in this phase, and in Sec.~\ref{spin-gapped phases} we analyze possible further instabilities in the spin sector and properties of the resulting phases.  We conclude with discussion of the orbital field effects in the context of the Spin Bose-Metal phase of Ref.~\onlinecite{Sheng09} where the Mott insulator is first produced by an eight-fermion Umklapp and the new four-fermion Umklapp appears as a residual interaction.

\section{Weak coupling approach to electrons on a zigzag strip with orbital field}\label{weakcoupling:orbital}

\begin{figure}[t]
  \includegraphics[width=\columnwidth]{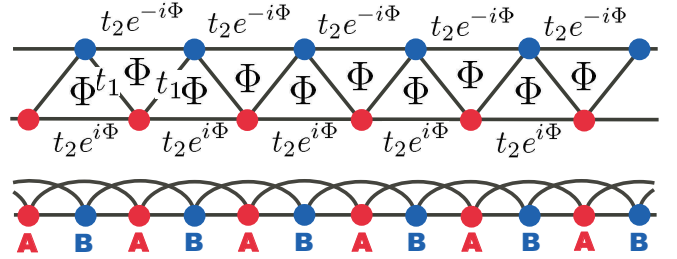}\\
  \caption{
Top:  Zigzag strip with uniform flux $\Phi$ penetrating each triangular plaquette.  Bottom:  Convenient representation of the model as a 1D chain with first- and second-neighbor hoppings.  We choose a gauge such that $t_{x,x+1} = t_1$ and $t_{x,x+2} = t_2 e^{i \Phi \cos{(\pi x)}}$.  The unit cell consists of two sites labelled ${\bf A}$ and ${\bf B}$.
}
  \label{strip with flux}
\end{figure}

\begin{figure}[t]
  \includegraphics[width=\columnwidth]{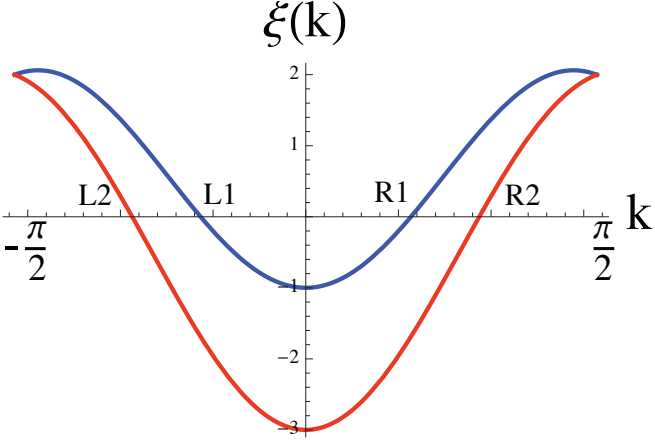}
  \caption{
Free electron spectrum in the presence of the orbital field, cf.~Fig.~\ref{strip with flux}.  Here $\xi(k)$ is given by Eq.~(\ref{xi_k}) with two branches and we focus on the regime when both bands are partially populated; we take $t_1=1$, $t_2=1$, and $\Phi=\pi/100$ for illustration.  The half-filling condition requires $k_{F1} + k_{F2} = \pi/2$.
}
  \label{orbdispersion}
\end{figure}

Let us apply weak coupling Renormalization Group (RG) to study effects of electronic interactions in the presence of the orbital magnetic field.  We start with free electrons hopping on the triangular strip with uniform flux $\Phi$ passing through each triangle.  Figure~\ref{strip with flux} illustrates our gauge choice,
\begin{eqnarray}
t_{x, x+1} &=& t_1 ~, \label{t1}\\
t_{x, x+2} &=& t_2 \, e^{i \Phi \cos{(\pi x)}} ~.\label{t2}
\end{eqnarray}
Here and throughout, we refer to sites by their 1D chain coordinate $x$.  Since the second-neighbor hopping depends on whether $x$ is even or odd, the unit cell has two sites which we label ${\bf A}$ and ${\bf B}$.  The Hamiltonian for such an interacting electron system is $H = H_0 + H_V$, with
\begin{eqnarray}
H_0 &=& -\sum_{x; \alpha} \left[ t_1 c^\dagger_\alpha(x) c_\alpha(x+1) + \Hc \right] \\
&& -\sum_{x \in {\bf A}; \alpha} \left[ t_2 e^{-i\Phi} c^\dagger_{{\bf A} \alpha}(x) \, c_{{\bf A} \alpha}(x+2) + \Hc \right] \\
&& -\sum_{x \in {\bf B}; \alpha} \left[ t_2 e^{i\Phi} c^\dagger_{{\bf B} \alpha}(x) \, c_{{\bf B} \alpha}(x+2) + \Hc \right] ~, \\
H_V &=& \frac{1}{2} \sum_{x,x'} V(x-x') n(x) n(x') ~.
\end{eqnarray}
In the first and last lines, we suppressed the sublattice labels, and $n(x) \equiv \sum_\alpha c_\alpha^\dagger(x) c_\alpha(x)$.  We assume that $H_V$ is small and treat it as a perturbation to $H_0$.  The free electron dispersion is
\begin{eqnarray}
\nonumber \xi(k) &=& \pm 2 \sqrt{ [t_1 \cos{(k)}]^2 + [t_2 \sin{(\Phi)} \sin{(2k)}]^2 } \\
&& -2 t_2 \cos{(\Phi)} \cos{(2k)} - \mu ~.
\label{xi_k}
\end{eqnarray}
We are focusing on the regime with two partially filled bands as shown in  Fig.~\ref{orbdispersion}.  For small flux, this regime appears when $t_2/t_1 > 0.5$.  We denote Fermi wavevectors for the right-moving electrons as $k_{F1}$ and $k_{F2}$ and the corresponding Fermi velocities as $v_1$ and $v_2$.  The half-filling condition reads $k_{F1} + k_{F2} = \pi/2$.

The electron operators are expanded in terms of continuum fields,
\begin{eqnarray}
c_{{\bf M} \alpha}(x) = \sum_{P,a} e^{i P k_{Fa} x} U^{\bf M}_{Pa} c_{Pa\alpha} ~,
\end{eqnarray}
where $P = R/L = +/-$ denotes the right and left movers, $a = 1, 2$ denotes the two Fermi seas, and ${\bf M} = {\bf A}$ or ${\bf B}$ denotes the sublattices.  In the specific gauge, the wavefunctions $U^{\bf M}_{Pa}$ are
\begin{equation}
\begin{array}{ll}
U^{{\bf A}}_{R1} = \cos{\left(\frac{\theta_{k_{F1}}}{2} \right)} ~, &~
U^{{\bf A}}_{L1} = \sin{\left(\frac{\theta_{k_{F1}}}{2} \right)} ~, \\
U^{{\bf A}}_{R2} = -\sin{\left(\frac{\theta_{k_{F2}}}{2} \right)} ~, &~
U^{{\bf A}}_{L2} =  \cos{\left(\frac{\theta_{k_{F2}}}{2} \right)} ~, \\
U^{{\bf B}}_{R1} = \sin{\left(\frac{\theta_{k_{F1}}}{2} \right)} ~, &~
U^{{\bf B}}_{L1} = \cos{\left(\frac{\theta_{k_{F1}}}{2} \right)} ~, \\
U^{{\bf B}}_{R2} =  \cos{\left(\frac{\theta_{k_{F2}}}{2} \right)} ~, &~
U^{{\bf B}}_{L2} = -\sin{\left(\frac{\theta_{k_{F2}}}{2} \right)} ~,
\end{array}
\end{equation} 
with
\begin{eqnarray}
\{\sin(\theta_k),\; \cos(\theta_k) \} \;\;\propto\;\; \{ t_1 \cos(k), \; t_2 \sin(\Phi) \sin(2k) \} .
\end{eqnarray}
Note that $k$ belongs to the reduced Brillouin zone $[-\pi/2, \pi/2]$.

Few words about physical symmetries.  The present problem has SU(2) spin rotation symmetry ($\mathcal{R}$) but lacks time reversal because of the orbital field.  It also lacks inversion symmetry and translation by one lattice spacing.  However, the system is invariant under combined transformations such as inversion plus complex conjugation ($I^* : x \to -x, ~i \to -i$) and translation by one lattice spacing plus complex conjugation ($T_1^*: x \to x+1, ~i \to -i$).   Table~\ref{tab:transfprops} lists transformation properties of the continuum fields under these two discrete transformations and under the SU(2) spin rotation.
Since the symmetries are reduced compared to the case without the orbital field,\cite{Lai10, Louis01, Fabrizio96} we need to scrutinize interactions allowed in the continuum field theory.

\begin{table}  
\caption{Transformation properties of the continuum fields under $I^*$ (inversion plus complex conjugation), $T_1^*$ (translation by one lattice spacing plus complex conjugation), and $\mathcal{R}$ (SU(2) spin rotation about arbitrary axis $\vec{n}$ by an angle $\phi$).  We also show transformation properties of bilinears $E_{1,2}$ defined in Eqs.~(\ref{E1})-(\ref{E2}).
}\label{tab:transfprops} 
\begin{ruledtabular}  
\begin{tabular}{c | c | c | c }  
  & $\mathcal{R}$ & $I^*$ & $T_1^*$ \\
  \hline
  $c_{Pa\alpha} \to$ &  $\left( e^{-i\frac{\phi}{2}\vec{n}\cdot \vec{\sigma}} \right)_{\alpha \beta} c_{Pa\beta}$ & $c_{Pa\alpha} $  & $e^{i P k_{Fa}} c_{-P,a\alpha}$ \\
  \hline
  $E_j \to$ & $E_j $        & $E_j$          & $-i E_j^\dagger$ \\
  \hline
  $E^{\dagger}_j \to$ & $E_j^\dagger $ & $E_j^\dagger$ & $i E_j$ 
\end{tabular}  
\end{ruledtabular}  
\end{table}

Using symmetry considerations, we can write down the general form of the four-fermion interactions which mix the right and left moving fields:
\begin{eqnarray}
\mathcal{H}^{\rho} &=& \sum_{a,b} \big{(} w^{\rho}_{ab} J_{Rab} J_{Lab} + \lambda^{\rho}_{ab} J_{Raa} J_{Lbb} \big{)} ~, \label{chargeint:orb}\\
\mathcal{H}^{\sigma} &=& -\sum_{a,b} \big{(} w^{\sigma}_{ab} \vec{J}_{Rab} \cdot \vec{J}_{Lab} + \lambda^{\sigma}_{ab} \vec{J}_{Raa} \cdot \vec{J}_{Lbb} \big{)} ~, \label{spinint:orb}\\
\nonumber
\mathcal{H}^u &=& u_4 \big{(} c^{\dagger}_{R2\uparrow} c^{\dagger}_{R2\downarrow} c_{L1\uparrow} c_{L1\downarrow} - c^{\dagger}_{L2\uparrow} c^{\dagger}_{L2\downarrow} c_{R1\uparrow} c_{R1\downarrow} \\ 
&& ~~~~~~~~ + \Hc \big{)} ~,\label{additionH:orb}
\end{eqnarray}
where we defined
\begin{eqnarray}\label{usual current op}
J_{Pab} &\equiv& c^{\dagger}_{P a \alpha} c_{P b \alpha} ~, \label{ccurrent:orbital}\\
\vec{J}_{Pab} &\equiv& \frac{1}{2} c^{\dagger}_{P a \alpha} \vec{\sigma}_{\alpha \beta} c_{P b \beta} ~.\label{scurrent:orbital}
\end{eqnarray}
Note that besides the familiar momentum-conserving four-fermion interactions $\mathcal{H}^\rho$ and $\mathcal{H}^\sigma$, there is also an Umklapp-type interaction $\mathcal{H}^u$.

Using the symmetries of the problem, we can check that all couplings are real and satisfy $w_{12} = w_{21}$ and $\lambda_{12} = \lambda_{21}$, and we also use convention $w_{11} = w_{22} = 0$.  Thus there are 9 independent couplings: $w^{\rho/\sigma}_{12}, \lambda^{\rho/\sigma}_{11}, \lambda^{\rho/\sigma}_{22}$, $\lambda^{\rho/\sigma}_{12}$, and $u_4$.

With all terms defined above, we can derive weak coupling RG equations:
\begin{eqnarray}
&&\hspace{-0.6cm}\dot{\lambda}^{\rho}_{11} = -\frac{1}{2\pi v_2} \left[ \left(w^{\rho}_{12}\right)^2+\frac{3}{16}\left(w^{\sigma}_{12}\right)^2\right],\label{rglambdarho11}\\
&&\hspace{-0.6cm}\dot{\lambda}^{\rho}_{22} = -\frac{1}{2\pi v_1} \left[ \left(w^{\rho}_{12}\right)^2+\frac{3}{16}\left(w^{\sigma}_{12}\right)^2\right],\label{rglambdarho22}\\
&&\hspace{-0.6cm}\dot{\lambda}^{\rho}_{12} = \frac{1}{\pi (v_1+v_2)} \left[ \left(w^{\rho}_{12}\right)^2+\frac{3}{16}\left(w^{\sigma}_{12}\right)^2 + \left( u_{4}\right)^2 \right],\label{rglambdarho12}\\
&&\hspace{-0.6cm}\dot{\lambda}^{\sigma}_{11} = -\frac{1}{2\pi v_1}\left(\lambda^{\sigma}_{11} \right)^2 -\frac{1}{4\pi v_2}\left[ \left( w^{\sigma}_{12}\right)^2+4w^{\rho}_{12}w^{\sigma}_{12}\right],\label{rglambdasigma11}\\
&&\hspace{-0.6cm}\dot{\lambda}^{\sigma}_{22} = -\frac{1}{2\pi v_2} \left(\lambda^{\sigma}_{22}\right)^2 -\frac{1}{4\pi v_1}\left[ \left(w^{\sigma}_{12}\right)^2+4w^{\rho}_{12}w^{\sigma}_{12}\right],\label{rglambdasigma22}\\
&&\hspace{-0.6cm} \dot{\lambda}^{\sigma}_{12} = -\frac{1}{\pi(v_1+v_2)}\left\{\left(\lambda^{\sigma}_{12}\right)^2+\frac{\left( w^{\sigma}_{12}\right)^2-4w^{\rho}_{12}w^{\sigma}_{12}}{2}\right\},\hspace{0.2cm}\label{rglambdasigma12}\\
&&\hspace{-0.6cm}\dot{w}^{\rho}_{12} = -\Lambda^{\rho}w^{\rho}_{12}-\frac{3}{16} \Lambda^{\sigma}w^{\sigma}_{12} ~,\label{rgwrho12}\\
&&\hspace{-0.6cm}\dot{w}^{\sigma}_{12} = -\Lambda^{\sigma}w^{\rho}_{12}-\left(\Lambda^{\rho}+\frac{\Lambda^{\sigma}}{2}+\frac{2 \lambda^{\sigma}_{12}}{\pi(v_1 + v_2)}\right) w^{\sigma}_{12} ~,\label{rgwsigma12}\\
&&\hspace{-0.6cm} \dot{u}_{4} = \frac{4 \lambda^{\rho}_{12} u_{4}}{\pi (v_1 + v_2)} ~.\label{u4}
\end{eqnarray}
Here $\dot{O} \equiv \partial{O}/\partial{\ell}$, where $\ell$ is logarithm of the length scale.  We have also defined
\begin{equation}
\Lambda^{\rho/\sigma} = \frac{\lambda^{\rho/\sigma}_{11}}{2\pi v_1}
+ \frac{\lambda^{\rho/\sigma}_{22}}{2\pi v_2}
- \frac{2\lambda^{\rho/\sigma}_{12}}{\pi (v_1 + v_2)} ~.
\end{equation}

We can obtain bare values of the couplings for any electronic interactions by expanding in terms of the continuum fields.  In the case of small flux, the couplings $\lambda^{\rho/\sigma}$ and $w^{\rho/\sigma}$ in Eqs.~(\ref{chargeint:orb}) and (\ref{spinint:orb}) are only modified slightly and can be treated as the same as in Ref.~\onlinecite{Lai10} with extended repulsion.  For the coupling $u_4$ in Eq.~(\ref{additionH:orb}), the bare value of $u_4$ in the small flux limit is $\sum_{x'} V(x-x') e^{i\frac{\pi}{2}(x-x')}\times \frac{t_2}{t_1}[\sin(k_{F1})+\sin(k_{F2})]\Phi \propto \Phi$, where $x$ and $x'$ belong to the same sublattice (${\bf A}$ or ${\bf B}$).  Therefore, we can see that the parameter $u_4$ which measures the strength of the umklapp process is linearly proportional to the flux and goes to zero if we gradually switch off the flux.  For repulsive interactions, we generally expect positive $\lambda^\rho$ (see, e.g., Ref.~\onlinecite{Lai10} with extended repulsion).  Then according to the RG Eq.~(\ref{u4}), positive initial $\lambda^\rho_{12}$ will drive $u_4$ to increase exponentially.  Thus we conclude that the starting two-band metallic phase is unstable due to the new Umklapp term.

To analyze the resulting phase(s), we use bosonization to rewrite fermionic fields in terms of bosonic fields,
\begin{equation}
c_{Pa\alpha} \sim \eta_{a\alpha} e^{i (\varphi_{a\alpha} + P\theta_{a\alpha})} ~,
\end{equation}
with canonically conjugate boson fields:
\begin{eqnarray} \label{commutation relation}
[\varphi_{a\alpha}(x) , \varphi_{b\beta}(x^\prime)] &=&
[\theta_{a\alpha}(x) , \theta_{b\beta}(x^\prime)] = 0 ~, \\ \relax
[\varphi_{a\alpha}(x) , \theta_{b\beta}(x^\prime)] &=&
i \pi \delta_{ab} \delta_{\alpha\beta} \, \Theta(x - x^\prime) ~,
\end{eqnarray}
where $\Theta(x)$ is the Heaviside step function.  Here we use Majorana fermions $\{ \eta_{a\alpha}, \eta_{b\beta}\} = 2\delta_{ab} \delta_{\alpha \beta}$ as Klein factors, which assure that the fermion fields with different flavors anticommute.

It is convenient to introduce new variables
\begin{eqnarray}
\theta_{\rho \pm} &\equiv& \frac{1}{2}[\theta_{1\up} + \theta_{1\dn} \pm (\theta_{2\up} + \theta_{2\dn})] ~, \label{rho+} \\
\theta_{a\sigma} &\equiv& \frac{1}{\sqrt{2}} (\theta_{a\up} - \theta_{a\dn}) ~, \hspace{0.5cm} a = 1~{\rm or}~2 ~, \label{asigma}\\
\theta_{\sigma \pm} &\equiv& \frac{1}{\sqrt{2}} (\theta_{1\sigma} \pm \theta_{2\sigma} ) ~, \label{sigma+}
\end{eqnarray}
and similarly for $\varphi$ variables.  We can then write compactly all nonlinear potentials obtained upon bosonization of the four-fermion interactions:
\begin{eqnarray}
&&\hspace{-0.5cm} \mathcal{H}^u = 4 u_4 \hat{\Gamma} \sin(2\varphi_{\rho-})\sin(2\theta_{\rho+}) ~,\label{u4boson}\\
&&\hspace{-0.5cm} W \equiv (w_{12}^\rho J_{R12} J_{L12} 
- w_{12}^\sigma \vec{J}_{R12} \cdot \vec{J}_{L12}) + \Hc = ~~~~~~~~ 
\label{w12} \\
\nonumber &&\hspace{-0.5cm} = \cos(2\varphi_{\rho-}) \Bigg\{
4 w_{12}^\rho \Big[\cos(2\varphi_{\sigma-}) 
                   - \hat{\Gamma} \cos(2\theta_{\sigma-}) \Big] ~~~~~~~ 
\label{w12rho} \\
&&\hspace{-0.5cm} - w_{12}^\sigma \Big[\cos(2\varphi_{\sigma-}) 
                       + \hat{\Gamma} \cos(2\theta_{\sigma-}) 
                       + 2 \hat{\Gamma} \cos(2\theta_{\sigma+}) \Big]
\Bigg\}, \label{w12sigma}\\
&&\hspace{-0.5cm} V_{\perp} \equiv -\sum_a \frac{\lambda^{\sigma}_{aa}}{2} \big{(} J^{+}_{Raa} J^{-}_{Laa} + J^{-}_{Raa} J^{+}_{Laa} \big{)} \\
&& \hspace{0.1cm} -\frac{\lambda^{\sigma}_{12}}{2} \big{[} J^{+}_{R11} J^{-}_{L22} + J^{-}_{R11} J^{+}_{L22} + (R \leftrightarrow L) \big{]} ~\\ 
&& = \sum_a \lambda^{\sigma}_{aa} \cos{(2\sqrt{2}\theta_{a\sigma})}\\
&& \hspace{0.1cm} + 2 \lambda^{\sigma}_{12} \hat{\Gamma} \cos{(2\theta_{\sigma+})} \cos{(2\varphi_{\sigma-})} ~,\label{vperp}
\end{eqnarray}
where
\begin{equation}
\hat{\Gamma} \equiv \eta_{1\up} \eta_{1\dn} \eta_{2\up} \eta_{2\dn} ~.
\end{equation}

We will not analyze the RG flows in all cases.  Our main interest is in exploring the orbital magnetic field effects on the C2S2 metallic phase and nearby C1$[\rho-]$S2 spin liquid.  Therefore we consider the situation where in the absence of the $u_4$ term we have the stable C2S2 phase described by RG flows such that $\lambda_{ab}^\rho$ reach some fixed point values, $w_{12}^{\rho/\sigma}$ are irrelevant, and $\lambda_{ab}^\sigma$ are marginally irrelevant -- this is realized, for example, in Ref.~\onlinecite{Lai10} for sufficiently long-ranged repulsion.  

As we have already discussed, for repulsive interactions we expect $\lambda_{12}^\rho > 0$ and hence any non-zero $u_4$ will increase quickly.  In this setting it is then natural to focus on the effects of the $\mathcal{H}^u$ first.  From the bosonized form Eq.~(\ref{u4boson}), we see that it pins
\begin{equation}\label{pinrho}
\sin(2\theta_{\rho+}) = -{\rm sign}(u_4) \sin(2\varphi_{\rho-}) = \pm 1 ~.
\end{equation}
Thus, both ``$\rho-$'' and ``$\rho+$'' modes become gapped and the system is an insulator.  This insulator arises because of the combined localizing effects of the orbital field and repulsive interactions.

Having concluded that $u_4$ becomes large, if we were to continue using the weak coupling RG Eqs.~(\ref{rglambdarho11})-(\ref{u4}), we would find that $u_4$ drives $\lambda_{12}^\rho$ to large positive value, which in turn drives $\Lambda^\rho$ to negative values and destabilizes couplings $w_{12}^{\rho/\sigma}$, and all couplings eventually diverge.  If we do not make finer distinctions as to which couplings diverge faster, we would conclude that the ultimate outcome is a fully gapped C0S0.  We will analyze different C0S0 phases arising from the combined effects of $u_4$ and $\lambda^\sigma$ later.  Here we only note that the bosonized theory suggests that a C0S2 phase can in principle be stable.  Indeed, once we pin $\varphi_{\rho-}$ to satisfy Eq.~(\ref{pinrho}), the $W$ interaction vanishes leaving only the effective $\lambda^\sigma$ couplings in the spin sector.  The stability in the spin sector is then determined by the signs of the $\lambda^\sigma$ couplings.  If $\lambda^\sigma_{ab} > 0$, the spin sector is stable and we have the C0S2 phase.  In what follows, we will identify all interesting physical observables in this phase and will use it as a starting point for analysis of possible further instabilities and features of the resulting phases.

\section{Observables in the Mott-insulating phase in orbital field}\label{observables}

To characterize the induced insulating phase(s), we consider observables constructed out of the fermion fields.  The only important bilinear operators are
\begin{eqnarray}
&& E_1 = \frac{1}{2} c^{\dagger}_{R1\alpha} c_{L2\alpha} + \frac{1}{2} c^{\dagger}_{R2\alpha} c_{L1\alpha} ~, \label{E1} \\
&& E_2 = \frac{1}{2} c^{\dagger}_{L2\alpha} c_{R1\alpha} - \frac{1}{2} c^{\dagger}_{L1\alpha} c_{R2\alpha} ~, \label{E2} \\
&& \vec{V}_1 = \frac{1}{2} c^{\dagger}_{R1\alpha} \vec{\sigma}_{\alpha\beta} c_{L2\beta} + \frac{1}{2} c^{\dagger}_{R2\alpha} \vec{\sigma}_{\alpha\beta} c_{L1\beta} ~,\\
&& \vec{V}_2 = \frac{1}{2} c^{\dagger}_{L2\alpha} \vec{\sigma}_{\alpha\beta} c_{R1\beta} - \frac{1}{2} c^{\dagger}_{L1\alpha} \vec{\sigma}_{\alpha\beta} c_{R2\beta} ~,
\end{eqnarray}
and their Hermitian conjugates.  All other bilinears contain field $\theta_{\rho-}$ and hence have exponentially decaying correlations once $\varphi_{\rho-}$ is pinned.  Here and below, repeated spin indices imply summation.  Operators $E_1, E_2$ are scalars and $\vec{V}_1, \vec{V}_2$ are vectors under spin SU(2).  One can check that $E_1$ and $E_2$ have identical transformation properties under all symmetries and therefore are not independent observables, and the same holds for $\vec{V}_1$ and $\vec{V}_2$.

The scalar bilinears $E_1$ and $E_2$ appear, e.g., when expressing fermion hopping energies and currents.  Specifically, consider a bond $[x, x'=x+n]$ (we will focus on $n=1$ or $2$),
\begin{eqnarray}
\mathcal{B}^{(n)}(x) \sim t_{x,x+n} c^{\dagger}_{\alpha}(x) c_{\alpha}(x+n) + \Hc ~, \\
\mathcal{J}^{(n)}(x) \sim i [t_{x,x+n} c^{\dagger}_{\alpha}(x) c_{\alpha}(x+n) - \Hc] ~,
\end{eqnarray}
where we have suppressed ``sublattice'' site labels ${\bf A}$ or ${\bf B}$ and $t_{x,x+n}$ is defined in Eqs.~(\ref{t1})-(\ref{t2}).  In general, we need to consider separately cases $[x\in {\bf A}, x' \in {\bf A}]$, $[x\in {\bf B}, x'\in {\bf B}]$, $[x\in {\bf A}, x'\in {\bf B}]$, $[x\in {\bf B}, x'\in {\bf A}]$.  After expansion in terms of the continuum fields in each case, we find that all cases can be summarized by a single form that requires only the physical coordinate $x$ but not the sublattice labels:
\begin{eqnarray}
\mathcal{B}^{(n)}(x)&:& e^{i \frac{\pi}{2} x} e^{i \frac{n}{2} \cdot \frac{\pi}{2}} \left( A^{(n)}_1 E^{\dagger}_1 + A^{(n)}_2 E^{\dagger}_2 \right) + \Hc ,\\
\mathcal{J}^{(n)}(x)&:& e^{i \frac{\pi}{2} x} e^{i \frac{n}{2} \cdot \frac{\pi}{2}} \left( A^{(n)}_3 E_1 + A^{(n)}_4 E_2 \right) + \Hc ,
\end{eqnarray}
where $A^{(n)}_{1,2,3,4}$ are some real numbers.  The above concise form is possible because of the $T_1^*$ symmetry involving translation by one lattice spacing.

In our analysis below, we will also use a scalar spin chirality defined as 
\begin{equation}\label{scalarspinchi}
\chi(x) = \vec{S}(x) \cdot[\vec{S}(x-1) \times \vec{S}(x+1)] ~.
\end{equation}
From the perspective of symmetry transformation properties, 
the scalar spin chirality and the so-called `site-centered' currents
\begin{equation}
\chi(x),~~~
\mathcal{J}^{(2)}(x-1),~~~
\mathcal{J}^{(1)}(x-1) + \mathcal{J}^{(1)}(x)
\end{equation}
have the same transformation properties. 
(Note that the above currents are named site-centered because they get inverted under inversion about site $x$.  Similarly, we can also call $\mathcal{J}^{(1)}(x)$ to be `bond-centered' since it is inverted under inversion about $x+1/2$, the center of the bond between $x$ and $x+1$.) 

Thus, up to some real factors, we can deduce that the scalar spin chirality in Eq.~(\ref{scalarspinchi}) contains the following contributions (focusing on terms that have power law correlations):
\begin{equation}
\chi(x) \;\;:\quad e^{i \frac{\pi}{2} x} \left( A^\prime_3 E_1 + A^\prime_4 E_2 \right) + \Hc ~.
\end{equation}

The vector bilinears $\vec{V}_1$ and $\vec{V}_2$ appear when expressing spin operator,
\begin{eqnarray}
\vec{S}(x) = \frac{1}{2} c^{\dagger}_\alpha(x) \vec{\sigma}_{\alpha\beta} c_\beta(x) ~.
\end{eqnarray}
We consider separately two cases $x\in {\bf A}$ and $x\in {\bf B}$.  After expanding in terms of the continuum fields, we find that both cases can be summarized by a single form that requires only the physical coordinate $x$,
\begin{eqnarray}
\vec{S}(x) \sim e^{i \frac{\pi}{2} x} \left( A'_1 \vec{V}^{\dagger}_1 + A'_2 \vec{V}^{\dagger}_2 \right) + \Hc~,
\end{eqnarray}
where $A'_{1,2}$ are some real factors. 

The bosonized expressions for $E_{1,2}$ are:
\begin{eqnarray}
\nonumber && E_1 = e^{-i\theta_{\rho+}} \Big[
-i \eta_{1\up} \eta_{2\up} e^{-i\theta_{\sigma+}}
\sin(\varphi_{\rho-} + \varphi_{\sigma-}) \\
&&\hspace{2.3cm}
-i \eta_{1\dn} \eta_{2\dn} e^{i\theta_{\sigma+}}
\sin(\varphi_{\rho-} - \varphi_{\sigma-})
\Big] ~,
\label{E1boson} \\
\nonumber && E_2 = e^{i\theta_{\rho+}} \Big[
\eta_{1\up} \eta_{2\up} e^{i\theta_{\sigma+}}
\cos(\varphi_{\rho-} + \varphi_{\sigma-}) ~~~~~~~~\\
&& \hspace{2cm}
+ \eta_{1\dn} \eta_{2\dn} e^{-i\theta_{\sigma+}}
\cos(\varphi_{\rho-} - \varphi_{\sigma-})
\Big] ~. ~~~~~~~~
\label{E2boson}
\end{eqnarray}
The bosonized expressions for $\vec{V}_1$ and $\vec{V}_2$ are similarly straightforward.  Since we have SU(2) spin invariance, for simplicity, we only write out $V^z$:
\begin{eqnarray}
\nonumber V^z_1 &=& e^{-i\theta_{\rho+}} \Big[ 
-i \eta_{1\up} \eta_{2\up} e^{-i \theta_{\sigma+}}
\sin(\varphi_{\rho-} + \varphi_{\sigma-})\\
&&~~~~~~~~~
+i \eta_{1\dn} \eta_{2\dn} e^{i \theta_{\sigma+}}
\sin(\varphi_{\rho-} - \varphi_{\sigma-})
\Big] ~, \\
\nonumber V^z_2 &=& e^{i\theta_{\rho+}} \Big[ 
\eta_{1\up} \eta_{2\up} e^{i \theta_{\sigma+}} \cos(\varphi_{\rho-} + \varphi_{\sigma-})\\
&&~~~~~~~~ 
- \eta_{1\dn} \eta_{2\dn} e^{-i \theta_{\sigma+}} \cos(\varphi_{\rho-} - \varphi_{\sigma-}) \Big] ~.
\end{eqnarray}

Besides the bilinears considered above, we have also identified important four-fermion operators,
\begin{eqnarray}
{\cal B}^{(1)}_{\rm stagg, I} &=& i (c_{R1}^\dagger \sigma^0 c_{L1}) (c_{R2}^\dagger \sigma^0 c_{L2})  + \Hc \nonumber \\
&\sim& \Big[\cos(2\theta_{\sigma+}) + \cos(2\theta_{\sigma-}) \Big] \sin(2\theta_{\rho+}) ~, \\
{\cal B}^{(1)}_{\rm stagg, II} &=& i (c_{R1}^\dagger \vec{\sigma} c_{L1}) \cdot (c_{R2}^\dagger \vec{\sigma} c_{L2}) + \Hc \nonumber \\
\nonumber &\sim& \Big[\cos(2\theta_{\sigma+}) - \cos(2\theta_{\sigma-}) + 2 \hat{\Gamma} \cos(2\varphi_{\sigma-}) \Big] \times \\
&& \times \sin(2\theta_{\rho+}) ~; \\
S^z_{\rm stagg, I} &=& (c_{R1}^\dagger \sigma^z c_{L1}) (c_{R2}^\dagger \sigma^0 c_{L2})  + \Hc \nonumber \\ 
&\sim& \Big[\sin(2\theta_{\sigma+}) + \sin(2\theta_{\sigma-}) \Big] \sin(2\theta_{\rho+}) ~, \\
S^z_{\rm stagg, II} &=& (c_{R1}^\dagger \sigma^0 c_{L1}) (c_{R2}^\dagger \sigma^z c_{L2})  + \Hc \nonumber \\
&\sim& \Big[\sin(2\theta_{\sigma+}) - \sin(2\theta_{\sigma-}) \Big] \sin(2\theta_{\rho+}) ~.
\end{eqnarray}
$\sigma^{0}$ above is the 2 $\times$ 2 identity matrix and $\vec{\sigma}$ are the usual Pauli matrices. The label ``staggered'' informs how they contribute to the spin and bond energy observables,
\begin{eqnarray}
\mathcal{B}^{(1)}(x)&:& e^{i \pi x} (A_{\rm I} \mathcal{B}^{(1)}_{\rm stagg, I} + A_{\rm II} \mathcal{B}^{(1)}_{\rm stagg, II}) ~, \\
S^z(x)&:& e^{i \pi x} (A'_{\rm I} S^z_{\rm stagg, I} + A'_{\rm II} S^z_{\rm stagg, II}) ~.
\end{eqnarray}
As an example, the above contributions to the bond energy arise from expanding nearest-neighbor energies $n(x) n(x+1)$ and $\vec{S}(x) \cdot \vec{S}(x+1)$ in terms of the continuum fields.  Again, we need to consider separately cases $x \in {\bf A}$ or $x \in {\bf B}$, but we find that both can be summarized by the form that requires only the physical coordinate $x$.

Note that we have only listed observables containing $\sin{(2\theta_{\rho+})}$.  Expressions that contain $\cos{(2\theta_{\rho+})}$ vanish because of the pinning condition Eq.~(\ref{pinrho}); in particular, there is no $\mathcal{B}^{({\rm n=even})}_{\rm stagg}$.  Also, for brevity we have only listed the bosonized form of the $z$-component of the spin observable.

There are several other non-vanishing four-fermion terms.  Thus, there is a term which can be interpreted as a staggered scalar spin chirality; however, it is identical to $\mathcal{H}^u$, Eq.~(\ref{additionH:orb}), and is always present as a static background in our system.  In addition, there is a spin-1 observable which can be interpreted as a spin current, and a spin-2 (i.e., spin-nematic) observable.  In the C0S2 phase, these will have the same power laws as $\mathcal{B}^{(1)}_{\rm stagg}$ and $\vec{S}_{\rm stagg}$.  However, in our model, they become short-ranged if any spin mode gets gapped, and we do not list them explicitly as the main observables.

Let us briefly describe treatment of the Klein factors (see, e.g., Ref.~\onlinecite{Fjaerestad02} for more details).  We need this in the next section when determining ``order parameters'' of various phases obtained as instabilities of the C0S2 phase.  The operator $\hat\Gamma = \eta_{1\up} \eta_{1\dn} \eta_{2\up} \eta_{2\dn}$ has eigenvalues $\pm 1$.  For concreteness, we work with the eigenstate corresponding to $+1$: $\hat\Gamma |+\ra = |+\ra$.  We then find the following relation
\begin{eqnarray}
\la +| \eta_{1\up} \eta_{2\up} |+\ra =
\la +| \eta_{1\dn} \eta_{2\dn} |+\ra = \textrm{pure imaginary} ~,
\end{eqnarray}
and the scalar bilinears are expressed as
\begin{eqnarray}
\label{epsilon_halfpi_fixGamma}
E_1 &\!=\!&
- e^{-i\theta_{\rho+}} \la +|\eta_{1\up} \eta_{2\up} |+\ra
\Big[
\cos(\varphi_{\rho-}) \sin(\theta_{\sigma+}) \sin(\varphi_{\sigma-})
\nonumber \\
&&~~~~~~~~~~~~~~~~~
+ i \sin(\varphi_{\rho-}) \cos(\theta_{\sigma+}) \cos(\varphi_{\sigma-})
\Big] ~, \\
\label{chi_halfpi_fixGamma}
E_2 &\!=\!&
e^{i\theta_{\rho+}} \la +|\eta_{1\up} \eta_{2\up} |+\ra
\Big[
\cos(\varphi_{\rho-}) \cos(\theta_{\sigma+}) \cos(\varphi_{\sigma-})
\nonumber \\
&&~~~~~~~~~~~~~~~~~
- i \sin(\varphi_{\rho-}) \sin(\theta_{\sigma+}) \sin(\varphi_{\sigma-})
\Big] ~.
\end{eqnarray}

For repulsively interacting electrons, the Umklapp term $\mathcal{H}^u$ appearing in the presence of the orbital field is always relevant and pins $\theta_{\rho+}$ and $\varphi_{\rho-}$ as in Eq.~(\ref{pinrho}).  As already discussed, for such pinning the $W$-term Eq.~(\ref{w12}) vanishes.  Therefore, as far as further instabilities of this C0S2 Mott insulator are concerned, we need to discuss the $V_\perp$-terms Eq.~(\ref{vperp}) that can gap out fields in the spin sector.

The instabilities depend on the signs of the couplings $\lambda^\sigma_{11}$, $\lambda^\sigma_{22}$, and $\lambda^\sigma_{12}$, so there are eight cases.  The simplest case is when all three $\lambda^{\sigma}_{ab} > 0$ and are all marginally irrelevant.  In this case, the phase is C0S2$[1\sigma, 2\sigma]$ with two gapless modes in the spin sector.  SU(2) spin invariance fixes the Luttinger parameters in the spin sector, $g_{1\sigma} = g_{2\sigma} = 1$.  After pinning the $\theta_{\rho+}$ and $\varphi_{\rho-}$, the scaling dimensions for the observables are
\begin{eqnarray}
\Delta[E_{1,2}] &=& \Delta[\vec{V}_{1,2}] = 1/2 ~,\\
\Delta[\mathcal{B}^{(1)}_{\rm stagg}]&=& \Delta[\vec{S}_{\rm stagg}] = 1 ~.
\end{eqnarray}
Thus we have spin and bond energy correlations oscillating with period 4 and decaying with power law $1/x$.

\section{Spin-gapped phases in orbital field}\label{spin-gapped phases}
Besides the spin-gapless phase, C0S2, there are other cases in which the spin sector is partially or fully gapped.  Below we discuss each case in detail and summarize the main properties in Table~\ref{tab:chi-relevant}.

\begin{table}
\begin{tabular}{|c|c|c|c|c|}
\hline
\multicolumn{1}{|c|}{$\lambda^\sigma_{11}$} &
\multicolumn{1}{|c|}{$\lambda^\sigma_{22}$} &
\multicolumn{1}{|c|}{$\lambda^\sigma_{12}$} &
\multicolumn{1}{|c|}{Static Order} &
\multicolumn{1}{|c|}{Power-Law Correlations}
\\
\hline
\multicolumn{1}{|c|}{\multirow{3}{*}{+}} &
\multicolumn{1}{|c|}{\multirow{3}{*}{+}} &
\multicolumn{1}{|c|}{\multirow{3}{*}{+}} &
\multicolumn{1}{|c|}{\multirow{3}{*}{None}} &
\multicolumn{1}{|c|}{$E_1$, $E_2$;}
 \\
\multicolumn{1}{|c|}{\multirow{3}{*}{}} &
\multicolumn{1}{|c|}{\multirow{3}{*}{}} &
\multicolumn{1}{|c|}{\multirow{3}{*}{}} &
\multicolumn{1}{|c|}{\multirow{3}{*}{}} &
\multicolumn{1}{|c|}{$\vec{V}_1$, $\vec{V}_2$;}
\\
\multicolumn{1}{|c|}{\multirow{3}{*}{}} &
\multicolumn{1}{|c|}{\multirow{3}{*}{}} &
\multicolumn{1}{|c|}{\multirow{3}{*}{}} &
\multicolumn{1}{|c|}{\multirow{3}{*}{}} &
\multicolumn{1}{|c|}{$\vec{S}_{stagg}$, $\mathcal{B}^{(1)}_{\rm stagg}$}
\\
\hline
\multicolumn{1}{|c|}{-} &
\multicolumn{1}{|c|}{+} &
\multicolumn{1}{|c|}{+} &
\multicolumn{1}{|c|}{None} &
\multicolumn{1}{|c|}{$\vec{S}_{\rm stagg}$, $\mathcal{B}^{(1)}_{\rm stagg}$}
\\
\hline
\multicolumn{1}{|c|}{+} &
\multicolumn{1}{|c|}{-} &
\multicolumn{1}{|c|}{+} &
\multicolumn{1}{|c|}{None} &
\multicolumn{1}{|c|}{$\vec{S}_{\rm stagg}$, $\mathcal{B}^{(1)}_{\rm stagg}$}
\\
\hline
\multicolumn{1}{|c|}{\multirow{2}{*}{+}} &
\multicolumn{1}{|c|}{\multirow{2}{*}{+}} &
\multicolumn{1}{|c|}{\multirow{2}{*}{-}} &
\multicolumn{1}{|c|}{$E_1$,~$E_2$;} &
\multicolumn{1}{|c|}{\multirow{2}{*}{None}}
\\
\multicolumn{1}{|c|}{\multirow{2}{*}{}} &
\multicolumn{1}{|c|}{\multirow{2}{*}{}} &
\multicolumn{1}{|c|}{\multirow{2}{*}{}} &
\multicolumn{1}{|c|}{$\mathcal{B}^{(1)}_{\rm stagg}$} &
\multicolumn{1}{|c|}{\multirow{2}{*}{}}
\\
\hline
\multicolumn{1}{|c|}{\multirow{1}{*}{-}} &
\multicolumn{1}{|c|}{\multirow{1}{*}{-}} &
\multicolumn{1}{|c|}{\multirow{1}{*}{+}} &
\multicolumn{1}{|c|}{$\mathcal{B}^{(1)}_{\rm stagg}$} &
\multicolumn{1}{|c|}{None}
\\
\hline
\multicolumn{1}{|c|}{\multirow{1}{*}{$\pm$}} &
\multicolumn{1}{|c|}{\multirow{1}{*}{$\mp$}} &
\multicolumn{1}{|c|}{\multirow{1}{*}{-}} &
\multicolumn{1}{|c|}{?} &
\multicolumn{1}{|c|}{?}
\\
\hline
\multicolumn{1}{|c|}{\multirow{1}{*}{-}} &
\multicolumn{1}{|c|}{\multirow{1}{*}{-}} &
\multicolumn{1}{|c|}{\multirow{1}{*}{-}} &
\multicolumn{1}{|c|}{?} &
\multicolumn{1}{|c|}{?}
\\
\hline
\end{tabular}
\caption{
Summary of the properties of the phases from different instabilities in the spin sector.
}
\label{tab:chi-relevant}
\end{table}

\subsection{$\lambda^\sigma_{11} < 0$, $\lambda^\sigma_{22} > 0$, $\lambda^\sigma_{12} > 0$}
In this case, only $\lambda^\sigma_{11}$ is relevant and flows to strong coupling.  We pin $\theta_{1\sigma}$ such that $\cos{(2\sqrt{2}\theta_{1\sigma})} = 1$ and the phase is C0S1$[2\sigma]$.  We have $S^z_{\rm stagg} \sim \sin(\sqrt{2}\theta_{2\sigma})$ and $B^{(1)}_{\rm stagg} \sim \cos(\sqrt{2}\theta_{2\sigma})$, so both show $1/x$ power law correlations.

\subsection{$\lambda^\sigma_{11} > 0$, $\lambda^\sigma_{22} < 0$, $\lambda^\sigma_{12} > 0$}
In this case, we pin $\theta_{2\sigma}$ such that $\cos{(2\sqrt{2}\theta_{2\sigma})} = 1$.  The phase is C0S1$[1\sigma]$ and is qualitatively similar to the previous case.

\subsection{$\lambda^\sigma_{11}, \lambda^\sigma_{22} > 0$, $\lambda^\sigma_{12} < 0$}
\begin{figure}[t]
  \includegraphics[width=\columnwidth]{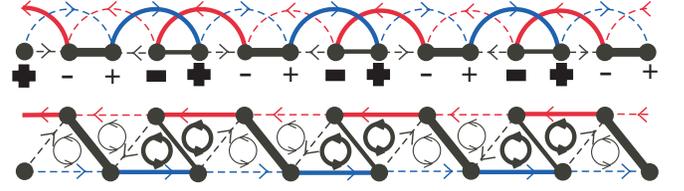}\\
  \caption{
Top:  Period 4 translational symmetry breaking when $\lambda^\sigma_{11}, \lambda^\sigma_{22} > 0$, $\lambda^\sigma_{12} < 0$, drawn in the 1D chain picture.  The bond energy pattern is given by Eqs.~(\ref{period4-bond1})-(\ref{period4-bond2}) and the chirality pattern by Eq. (\ref{period4-chi}).  Thicker lines represent stronger bond; ``+'' and ``-'' symbols of varying boldness schematize the scalar chirality associated with sites (or equivalently with site-centered loops); and arrows on the links show the bond currents.  Bottom: The same pattern in the two-leg triangular ladder drawing.
}
  \label{period4bond+chi}
\end{figure}

In this case, $\lambda^\sigma_{11}$ and $\lambda^\sigma_{22}$ are marginally irrelevant while $\lambda^\sigma_{12}$ is marginally relevant and flows to strong coupling.  To minimize the energy associated with $\lambda^\sigma_{12}$, cf.~Eq.~(\ref{vperp}), we pin $\theta_{\sigma+}$ and $\varphi_{\sigma-}$ to satisfy,
\begin{equation}
\cos{(2\theta_{\sigma+})} \cos{(2\varphi_{\sigma-})} = 1 ~.
\end{equation}
To characterize the resulting C0S0 fully gapped phase, we note that $E_j$ and $\mathcal{B}^{(1)}_{\rm stagg}$ gain expectation values.  We calculate the first- and second-neighbor bond energies,
\begin{eqnarray}
\nonumber \delta \mathcal{B}^{(1)}(x) &\sim& e^{i \frac{\pi}{2} x} e^{i\frac{\pi}{4}} \left( A^{(1)}_1 E^{\dagger}_1 + A^{(1)}_2 E^{\dagger}_2 \right) + \Hc \\
\nonumber && +\; e^{i \pi x} \mathcal{B}^{(1)}_{\rm stagg} \\
&\simeq& \tilde{A} \cos{\left(\frac{\pi}{2} x + \frac{\pi}{4} + \alpha \right)} + \tilde{C} \cos{(\pi x)} ~, \label{period4-bond1} \\
\nonumber \delta \mathcal{B}^{(2)}(x) &\sim& e^{i \frac{\pi}{2} x} e^{i\frac{\pi}{2}} \left( A^{(2)}_1 E^{\dagger}_1 + A^{(2)}_2 E^{\dagger}_2 \right) + \Hc \\
&\simeq& \tilde{A}' \cos{\left( \frac{\pi}{2} x + \frac{\pi}{2} + \alpha \right)} ~, \label{period4-bond2}
\end{eqnarray}
where $\tilde{A}$, $\tilde{C}$, and $\tilde{A}'$ are some non-universal real numbers, while $\alpha$ is fixed to one of the values $\{ \pm \frac{\pi}{4},~\pm \frac{3\pi}{4} \}$.  We see that this phase has translation symmetry breaking with period 4 as illustrated in Fig.~\ref{period4bond+chi}.  The four independent values of $\alpha$ correspond to four translations of the bond pattern along $x$.

To further characterize the state, we also calculate the scalar chirality,
\begin{eqnarray}
\chi(x) \sim \tilde{A}'' \cos{\left( \frac{\pi}{2} x - \frac{\pi}{2} + \alpha \right)} + \tilde{C}'' \cos{(\pi x)} ~,\label{period4-chi}
\end{eqnarray}
where $\tilde{A}''$ and $\tilde{C}''$ are some non-universal real amplitudes, while $\alpha$ is the same as in Eqs.~(\ref{period4-bond1})-(\ref{period4-bond2}).  The period-4 pattern induced in the chirality is also shown in Fig.~\ref{period4bond+chi} and is consistent with the spontaneous period-4 bond order on top of the staggered chirality background present from the outset.

\subsection{$\lambda^\sigma_{11}, \lambda^\sigma_{22} < 0$, $\lambda^\sigma_{12} > 0$}
\begin{figure}[t]
   \centering
   \includegraphics[width=\columnwidth]{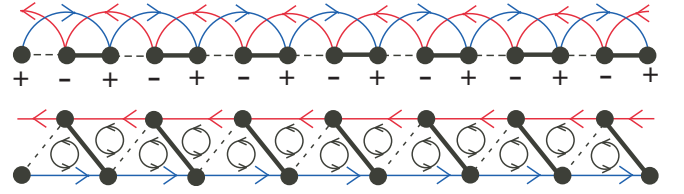}
   \caption{
Top: Static period-2 VBS when $\lambda^\sigma_{11}, \lambda^\sigma_{22} < 0$, $\lambda^\sigma_{12} > 0$, drawn in the 1D chain picture.  Note that the background staggered chirality is present from the outset due to the orbital field.  Bottom: The same pattern in the two-leg triangular ladder drawing.
}
   \label{p2vbschirality}
\end{figure}

In this case, $\lambda^\sigma_{11}$ and $\lambda^\sigma_{22}$ are marginally relevant and flow to strong coupling while $\lambda^\sigma_{12}$ is marginally irrelevant.  To minimize the relevant interactions, we pin
\begin{equation}
\cos{(2\sqrt{2}\theta_{1\sigma})} = \cos{(2\sqrt{2}\theta_{2\sigma})}=1 ~.
\end{equation}
This is a different C0S0 fully gapped phase where only $B^{(n={\rm odd})}_{\rm stagg}$ gain expectation values.  The nearest-neighbor bond energy is
\begin{eqnarray}
\delta\mathcal{B}^{(1)}(x) \simeq && e^{i\pi x} \mathcal{B}^{(1)}_{\rm stagg} = \tilde{C} \cos(\pi x) ~.
\end{eqnarray}
The physical picture of this phase is shown in Fig.~\ref{p2vbschirality}.

\subsection{$\lambda^\sigma_{12} < 0$ and either $\lambda^\sigma_{11} < 0$ or $\lambda^\sigma_{22} < 0$}
Here, we do not know how to minimize the relevant interactions due to the competition of the pinning conditions in $V_\perp$, Eq.~(\ref{vperp}).  However, we expect that, depending which terms grow faster under the RG and win, the final outcome reduces to one of the phases discussed above.

\section{Discussion}\label{conclusion}
In this paper, we considered the effects of orbital field on the half-filled electronic two-leg triangular ladder.  In weak coupling, the Umklapp $\mathcal{H}^u$ [Eqs.~(\ref{additionH:orb})~and ~(\ref{u4boson})] always makes the system Mott-insulating, and we described in detail possible phases.

We would like to conclude by indicating a connection with the Spin Bose-Metal (SBM) theory in Ref.~\onlinecite{Sheng09} and discussing effects of the orbital field on the SBM.  It turns out that our present electronic results translate readily to this case.  The SBM can be viewed as an intermediate coupling C1[$\rho-$]S2 phase and is obtained in the absence of the field by gapping out the overall charge mode using an {\it eight-fermion} Umklapp term, whose bosonized form is\cite{Sheng09}
\begin{equation}
H_8 = 2 v_8 \cos(4\theta_{\rho+}) ~.
\end{equation}
Reference~\onlinecite{Sheng09} argued that $v_8 > 0$ is appropriate for the electronic case that corresponds to a spin-1/2 system with ring exchanges on the zigzag ladder.  This gives pinning condition for the overall charge mode,
\begin{eqnarray}
4\theta_{\rho+} = \pi~~({\rm mod}~2\pi) ~.
\end{eqnarray}
Note that this pinning condition is compatible with the pinning Eq.~(\ref{pinrho}) due to the new four-fermion Umklapp $\mathcal{H}^u$ arising in the presence of the orbital field, so the two Umklapps lead to similar Mott insulators.

We can consider situations where the main driving force to produce Mott insulator is the eight-fermion Umklapp while the orbital field is a small perturbation onto the SBM phase.  Formulated entirely in the spin language, the underlying electronic orbital fields give rise to new terms in the Hamiltonian of a form $\vec{S}_1 \cdot [\vec{S}_2 \times \vec{S}_3]$ on each triangle circled in the same direction.\cite{Rokhsar90, Sen95, magnetoorb}  In the 1D chain language, this becomes a staggered spin chirality term $(-1)^x \vec{S}(x-1) \cdot [\vec{S}(x) \times \vec{S}(x+1)]$.  Starting from the SBM theory in the absence of the field, this gives a new residual interaction of the same form as $\mathcal{H}^u$ (similar to $\chi_\pi$ in Ref.~\onlinecite{Sheng09}).  In principle, this $\mathcal{H}^u$ can be irrelevant in the SBM phase if the one Luttinger parameter $g_{\rho-}$ in the SBM theory\cite{Sheng09} is less than 1/2, and in this case the orbital effects will renormalize down on long length scales.  On the other hand, if this terms is relevant and pins $\varphi_{\rho-}$, then the resulting phases are precisely as already considered in the electronic language.  In this simple-minded approach, all the phases we discussed in the paper are proximate to the SBM phase.  It would be interesting to explore spin models realizing the SBM in the presence of such additional chirality terms.\cite{Sheng09, YiFei09}

The presented physics appears to be rather special to the 2-leg ladder case, but is quite interesting in the context of such models.  Perhaps the most intriguing finding is the C0S2 phase with two gapless spin modes.  Note that the relevant chirality interaction involves both chains and the system is far from the regime of decoupled chains.  Our characterization of this state comes from the formal bosonization treatment, but it would be interesting to develop a simpler intuitive picture.

\section{acknowledgments}
We would like to thank M.~P.~A.~Fisher for useful discussions.  This research is supported by the National Science Foundation through grant DMR-0907145 and by the A.~P.~Sloan Foundation.

\bibliography{biblio4zigzag}

\end{document}